\documentclass[11pt]{article}

\usepackage[margin=1.1in]{geometry}
\usepackage[T1]{fontenc}
\usepackage[utf8]{inputenc}
\usepackage{lmodern}
\usepackage{microtype}
\usepackage{amsmath,amssymb,mathtools}
\usepackage{amsthm}
\usepackage{graphicx}
\usepackage{float}
\usepackage{xcolor}
\usepackage[round,authoryear]{natbib}
\usepackage[colorlinks=true,linkcolor=blue!45!black,citecolor=blue!45!black,urlcolor=blue!45!black]{hyperref}
\usepackage[nameinlink,capitalize]{cleveref}
\usepackage{pras}
\usepackage[most]{tcolorbox}
\usepackage{enumitem}
\usetikzlibrary{decorations.pathreplacing}
\setcitestyle{authoryear,round}

\graphicspath{{./}}

\DeclareMathOperator{\rank}{rank}
\newcommand{\dif}{\mathop{}\!\mathrm{d}}

\allowdisplaybreaks

\tcbuselibrary{theorems}

\definecolor{QuestionBlue}{HTML}{1B4D89}
\definecolor{PropGreen}{HTML}{2D6A4F}
\definecolor{ThmGreen}{HTML}{2D6A4F}
\definecolor{LemmaPurple}{HTML}{5D3A9B}
\definecolor{ConjOrange}{HTML}{B46000}

\tcbset{
  note box/.style={
    enhanced,
    breakable,
    sharp corners=south,
    boxrule=0.6pt,
    left=8pt,
    right=8pt,
    top=8pt,
    bottom=8pt,
    before skip=12pt,
    after skip=12pt,
    fonttitle=\bfseries,
    coltitle=black,
  }
}

\newtcbtheorem[
  auto counter,
  crefname={question}{questions},
  Crefname={Question}{Questions}
]{questionbox}{Question}{
  note box,
  separator sign={\ },
  colback=QuestionBlue!4,
  colframe=QuestionBlue,
  colbacktitle=QuestionBlue!65,
}{q}

\newtcbtheorem[
  auto counter,
  crefname={proposition}{propositions},
  Crefname={Proposition}{Propositions}
]{propbox}{Proposition}{
  note box,
  separator sign={\ },
  colback=PropGreen!4,
  colframe=PropGreen!85!black,
  colbacktitle=PropGreen!65,
}{prop}

\newtcbtheorem[
  auto counter,
  crefname={theorem}{theorems},
  Crefname={Theorem}{Theorems}
]{theorembox}{Theorem}{
  note box,
  separator sign={\ },
  colback=ThmGreen!4,
  colframe=ThmGreen!85!black,
  colbacktitle=ThmGreen!65,
}{thm}

\newtcbtheorem[
  auto counter,
  crefname={lemma}{lemmas},
  Crefname={Lemma}{Lemmas}
]{lemmabox}{Lemma}{
  note box,
  separator sign={\ },
  colback=LemmaPurple!4,
  colframe=LemmaPurple!85!black,
  colbacktitle=LemmaPurple!65,
}{lem}

\newtcbtheorem[
  auto counter,
  crefname={conjecture}{conjectures},
  Crefname={Conjecture}{Conjectures}
]{conjecturebox}{Conjecture}{
  note box,
  separator sign={\ },
  colback=ConjOrange!4,
  colframe=ConjOrange!90!black,
  colbacktitle=ConjOrange!65,
}{conj}

\def\comments{1}

\ifnum\comments=1
\newcommand{\pras}[1]{\textcolor{BrickRed}{\sf{[#1 --PR]}}}
\newcommand{\kangning}[1]{\textcolor{orange}{\sf{[#1 --KW]}}}
\newcommand{\moses}[1]{\textcolor{darkgreen}{\sf{[#1 --MC]}}}

\fi

\ifnum\comments=0
\newcommand{\pras}[1]{}
\newcommand{\kangning}[1]{}
\newcommand{\moses}[1]{}

\fi

\title{An Exposition of \\Five Candidates Suffice for a Majority}

\author{%
  Moses Charikar\footnotemark[1]
  \quad
  Prasanna Ramakrishnan\footnotemark[1]
  \quad
  Kangning Wang\footnotemark[2]
}
\date{\today}

\makeatletter
\renewcommand{\maketitle}{%
  \begingroup
  \renewcommand{\thefootnote}{\fnsymbol{footnote}}%
  \begin{center}
    {\LARGE\bfseries \@title\par}
    \vspace{0.35em}
    {\large \@author\par}
    \vspace{0.35em}
    {\large \@date\par}
  \end{center}
  \footnotetext[1]{Stanford University}
  \footnotetext[2]{Rutgers University}
  \endgroup
  \vspace{1.5em}
}
\makeatother

\begin{document}
\maketitle

\vspace{-3em}

\begin{abstract}
We give a brief exposition of a result of \cite*{DBLP:conf/soda/SongNL26} that every election (with ranked preferences) has a Condorcet winning set of at most five candidates.
\end{abstract}

\section{Problem statement, notation, and proof overview}

An \textit{election} consists of a set $V$ of $n$ voters, a set $C$ of $m$ candidates, and a linear preference order $\succ_v$ over the candidates for each voter $v \in V$. We use $\tfrac1n|a \succ S|$ to denote the fraction of voters that prefer candidate $a$ over each candidate in the committee (set of candidates) $S$. A \textit{Condorcet winning set} (introduced by \cite*{elkind2011choosing,elkind2015condorcet}) is a committee $S$ such that for all candidates $a \in C$, $\tfrac1n|a \succ S| < \frac12$.

The goal of this note is to prove the following result, due to \cite*{DBLP:conf/soda/SongNL26}. 

\begin{theorembox}{\citep*{DBLP:conf/soda/SongNL26}}{five}
In every election, there is a Condorcet winning set of size at most $5$.
\end{theorembox}

The general proof strategy used by \cite*{DBLP:conf/soda/SongNL26}, as well as its antecedents \cite*{jiang2020approximately} and \cite{DBLP:conf/stoc/CharikarLRV025}, is the following:

\begin{enumerate}[label=(\arabic*)]
    \item  Construct a distribution $D$ over candidates such that all candidates $a$ tend to be ``ranked low'' by voters (in comparison to $D$).
    
    \item Show that if we construct the committee $S$ using samples from $D$, with positive probability $S$ tends to be ``ranked high'' by voters (in comparison to $D$).
    
    \item Argue that if $a$ tends to be ranked low and $S$ tends to be ranked high, then a majority of voters cannot prefer $a$ over $S$.
\end{enumerate}

\cite*{jiang2020approximately} and \cite{DBLP:conf/stoc/CharikarLRV025} use distributions adapted from a \emph{stable lottery}  \citep{DBLP:journals/teco/ChengJMW20}, the equilibrium distribution of a certain two-player game, whose existence is proved via the minimax theorem. \cite*{DBLP:conf/soda/SongNL26} use a distribution adapted from a \textit{Lindahl equilibrium}, whose existence is proved via the Kakutani fixed-point theorem. At a very high level, the Lindahl equilibrium is a market equilibrium which gives each voter a fixed budget and  personalized prices for each candidate, and assigns the voters a common fractional allocation of candidates that each voter likes as much as what they would get if they could spend their budget freely. (We note that usually the agents in a Lindahl equilibrium have \textit{cardinal utilities}, and some effort is needed to adapt to a setting with ordinal preferences.)

While the market setup of the Lindahl equilibrium provides helpful intuition for why the distribution $D$ is reasonable, it can take some effort to parse for readers unfamiliar with market equilibria. In an effort to make the proof more accessible to this audience, we will present a more streamlined version of the \cite*{DBLP:conf/soda/SongNL26} proof which applies the Kakutani fixed-point theorem directly rather than going through the Lindahl equilibrium. An upcoming paper of \cite{nguyen2026stable} also uses a similar approach to give a proof of \Cref{thm:five} and several other results, and we emphasize that this note is purely for expository purposes.

As a final piece of notation, we formalize what it means for candidates to be ``ranked low'' or ``ranked high'' by voters. Voter $v$'s \emph{rank} of a candidate $a$ with respect to a distribution $D$ over candidates is the probability that $v$ prefers $a$ over a random candidate $b\sim D$. Borrowing notation from \cite{DBLP:conf/stoc/CharikarLRV025},
$$\rank_v(a; D) := \Pr_{b\sim D}[a \succ_v b].$$

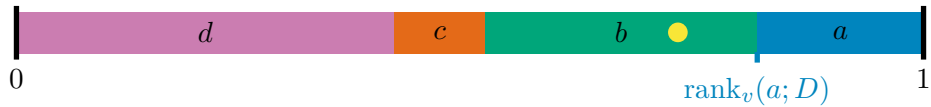
\begin{figure}[ht]
\centering
\begin{tikzpicture}[x=1cm,y=1cm]
  \definecolor{dcol}{RGB}{203,125,177}
  \definecolor{ccol}{RGB}{228,106,24}
  \definecolor{bcol}{RGB}{0,166,122}
  \definecolor{acol}{RGB}{0,133,190}
  \definecolor{dotcol}{RGB}{248,230,55}

  \def\H{0.55}
  \def\xD{0}
  \def\xC{5.0}
  \def\xB{6.2}
  \def\xA{9.8}
  \def\xR{12.0}

  \fill[dcol] (\xD,0) rectangle (\xC,\H);
  \fill[ccol] (\xC,0) rectangle (\xB,\H);
  \fill[bcol] (\xB,0) rectangle (\xA,\H);
  \fill[acol] (\xA,0) rectangle (\xR,\H);

  \draw[black,line width=2pt] (\xD,-0.08) -- (\xD,\H+0.08);
  \draw[black,line width=2pt] (\xR,-0.08) -- (\xR,\H+0.08);

  \node[below] at (\xD,-0.08) {$0$};
  \node[below] at (\xR,-0.08) {$1$};

  \node at ({(\xD+\xC)/2},0.28) {$d$};
  \node at ({(\xC+\xB)/2},0.28) {$c$};
  \node at ({(\xB+\xA)/2},0.28) {$b$};
  \node at ({(\xA+\xR)/2},0.28) {$a$};

  \fill[dotcol] (8.75,0.28) circle[radius=0.13];

  \draw[acol,line width=2pt] (\xA,-0.12) -- (\xA,0.02);
  \node[acol,below] at (\xA,-0.16) {$\rank_v(a;D)$};
\end{tikzpicture}
\caption{Depiction of the ranks for a voter $v$ with preference $a\succ b\succ c \succ d$ with respect to a distribution $D$ which chooses $(a, b, c, d)$ with probabilities $(0.2,0.3,0.1, 0.4)$. The yellow dot represents $r\sim \Unif(0, 1)$, which can be used to sample from $D$.}
\label{fig:rank-v-a-D}
\end{figure}

The intuition for the ranks is best understood with the following picture. Imagine creating a block for each candidate $a$, whose width is the probability mass of $a$ in $D$, and having each voter arrange these blocks along the interval $[0, 1]$ in increasing order of preference (see \Cref{fig:rank-v-a-D} for an example). Then $\rank_v(a;D)$ is precisely the bottom (leftmost point) of the block corresponding to $a$.

The pictorial representation also gives a helpful alternative way of understanding  samples from $D$. From each voter's perspective, $b\sim D$ is equivalent to sampling a real number $r\sim \Unif(0, 1)$, and choosing the candidate whose block contains $r$. (Formally, the candidate $b$ for which $\rank_v(b;D)$ is maximal subject to being at most $r$.) With this view we get the following two immediate consequences:
\begin{itemize}
    \item  The random variable $\rank_v(b;D)$ with $b \sim D$ is stochastically dominated by $r\sim \Unif(0, 1)$. 
    \item The random variable $\displaystyle\min_{a: a\succ_v b} \rank_v(a;D)$ with $b \sim D$ stochastically dominates $r\sim \Unif(0, 1)$.\footnote{If $b$ is $v$'s favorite candidate, then we define $\displaystyle\min_{a: a\succ_v b} \rank_v(a;D) = 1$ so that it is always the top (rightmost point) of the block corresponding to $b$.}
\end{itemize}

\section{Five candidates suffice}

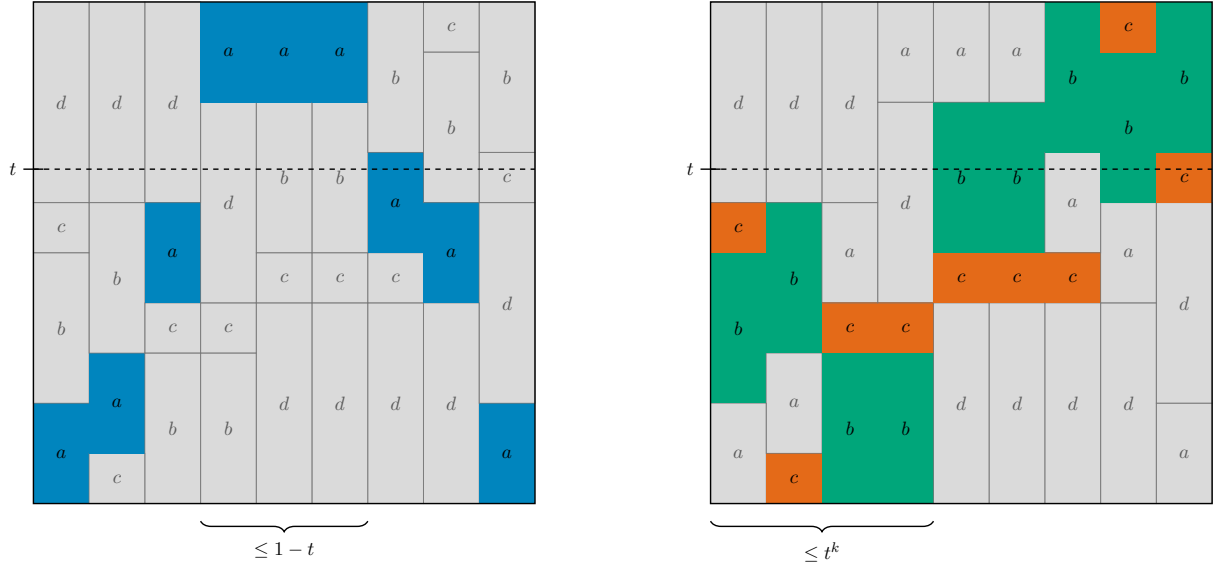
\begin{figure}[ht]
\centering
\resizebox{\textwidth}{!}{%
\begin{tikzpicture}[x=10cm,y=10cm]

  \definecolor{dcol}{RGB}{203,125,177}
  \definecolor{ccol}{RGB}{228,106,24}
  \definecolor{bcol}{RGB}{0,166,122}
  \definecolor{acol}{RGB}{0,133,190}
  \definecolor{lightgrayblock}{RGB}{150,150,150}
  \definecolor{grayborder}{RGB}{120,120,120}
  \definecolor{darkgraylabel}{RGB}{90,90,90}

  \def\hc{0.1}
  \def\ha{0.2}
  \def\hb{0.3}
  \def\hd{0.4}

  \pgfmathsetmacro{\W}{1/9}

  \def\sep{0.35}

  \newcommand{\Ablock}[2]{%
    \fill[acol] ({#1*\W},#2) rectangle ++(\W,\ha);
    \node at ({#1*\W + 0.5*\W},{#2 + 0.5*\ha}) {$a$};
  }

  \newcommand{\GrayBlock}[4]{%
    \filldraw[
      fill=lightgrayblock,
      fill opacity=0.35,
      draw=grayborder,
      draw opacity=0.8,
      line width=0.4pt
    ] ({#1*\W},#2) rectangle ++(\W,#3);
    \node[text=darkgraylabel] at ({#1*\W + 0.5*\W},{#2 + 0.5*#3}) {$#4$};
  }

  \newcommand{\ColorBlock}[5]{%
    \fill[#5] ({#1*\W},#2) rectangle ++(\W,#3);
    \node at ({#1*\W + 0.5*\W},{#2 + 0.5*#3}) {$#4$};
  }

  \begin{scope}[shift={(0,0)}]

    \Ablock   {0}{0.0}
    \GrayBlock{0}{0.2}{\hb}{b}
    \GrayBlock{0}{0.5}{\hc}{c}
    \GrayBlock{0}{0.6}{\hd}{d}

    \GrayBlock{1}{0.0}{\hc}{c}
    \Ablock   {1}{0.1}
    \GrayBlock{1}{0.3}{\hb}{b}
    \GrayBlock{1}{0.6}{\hd}{d}

    \GrayBlock{2}{0.0}{\hb}{b}
    \GrayBlock{2}{0.3}{\hc}{c}
    \Ablock   {2}{0.4}
    \GrayBlock{2}{0.6}{\hd}{d}

    \GrayBlock{3}{0.0}{\hb}{b}
    \GrayBlock{3}{0.3}{\hc}{c}
    \GrayBlock{3}{0.4}{\hd}{d}
    \Ablock   {3}{0.8}

    \GrayBlock{4}{0.0}{\hd}{d}
    \GrayBlock{4}{0.4}{\hc}{c}
    \GrayBlock{4}{0.5}{\hb}{b}
    \Ablock   {4}{0.8}

    \GrayBlock{5}{0.0}{\hd}{d}
    \GrayBlock{5}{0.4}{\hc}{c}
    \GrayBlock{5}{0.5}{\hb}{b}
    \Ablock   {5}{0.8}

    \GrayBlock{6}{0.0}{\hd}{d}
    \GrayBlock{6}{0.4}{\hc}{c}
    \Ablock   {6}{0.5}
    \GrayBlock{6}{0.7}{\hb}{b}

    \GrayBlock{7}{0.0}{\hd}{d}
    \Ablock   {7}{0.4}
    \GrayBlock{7}{0.6}{\hb}{b}
    \GrayBlock{7}{0.9}{\hc}{c}

    \Ablock   {8}{0.0}
    \GrayBlock{8}{0.2}{\hd}{d}
    \GrayBlock{8}{0.6}{\hc}{c}
    \GrayBlock{8}{0.7}{\hb}{b}

    \draw[black,dashed,thick] (0,{2/3}) -- (1,{2/3});

    \draw[black,thick] (-0.02,{2/3}) -- (0.02,{2/3});
    \node[left] at (-0.02,{2/3}) {$t$};

    \draw[black,thick] (0,0) rectangle (1,1);

    \draw[decorate,decoration={brace,mirror,amplitude=6pt},thick]
      ({3*\W},-0.035) -- ({6*\W},-0.035)
      node[midway,below=8pt] {$\leq 1-t$};

  \end{scope}

  \begin{scope}[shift={({1+\sep},0)}]

    \GrayBlock {0}{0.0}{\ha}{a}
    \ColorBlock{0}{0.2}{\hb}{b}{bcol}
    \ColorBlock{0}{0.5}{\hc}{c}{ccol}
    \GrayBlock {0}{0.6}{\hd}{d}

    \ColorBlock{1}{0.0}{\hc}{c}{ccol}
    \GrayBlock {1}{0.1}{\ha}{a}
    \ColorBlock{1}{0.3}{\hb}{b}{bcol}
    \GrayBlock {1}{0.6}{\hd}{d}

    \ColorBlock{2}{0.0}{\hb}{b}{bcol}
    \ColorBlock{2}{0.3}{\hc}{c}{ccol}
    \GrayBlock {2}{0.4}{\ha}{a}
    \GrayBlock {2}{0.6}{\hd}{d}

    \ColorBlock{3}{0.0}{\hb}{b}{bcol}
    \ColorBlock{3}{0.3}{\hc}{c}{ccol}
    \GrayBlock {3}{0.4}{\hd}{d}
    \GrayBlock {3}{0.8}{\ha}{a}

    \GrayBlock {4}{0.0}{\hd}{d}
    \ColorBlock{4}{0.4}{\hc}{c}{ccol}
    \ColorBlock{4}{0.5}{\hb}{b}{bcol}
    \GrayBlock {4}{0.8}{\ha}{a}

    \GrayBlock {5}{0.0}{\hd}{d}
    \ColorBlock{5}{0.4}{\hc}{c}{ccol}
    \ColorBlock{5}{0.5}{\hb}{b}{bcol}
    \GrayBlock {5}{0.8}{\ha}{a}

    \GrayBlock {6}{0.0}{\hd}{d}
    \ColorBlock{6}{0.4}{\hc}{c}{ccol}
    \GrayBlock {6}{0.5}{\ha}{a}
    \ColorBlock{6}{0.7}{\hb}{b}{bcol}

    \GrayBlock {7}{0.0}{\hd}{d}
    \GrayBlock {7}{0.4}{\ha}{a}
    \ColorBlock{7}{0.6}{\hb}{b}{bcol}
    \ColorBlock{7}{0.9}{\hc}{c}{ccol}

    \GrayBlock {8}{0.0}{\ha}{a}
    \GrayBlock {8}{0.2}{\hd}{d}
    \ColorBlock{8}{0.6}{\hc}{c}{ccol}
    \ColorBlock{8}{0.7}{\hb}{b}{bcol}

    \draw[black,dashed,thick] (0,{2/3}) -- (1,{2/3});

    \draw[black,thick] (-0.02,{2/3}) -- (0.02,{2/3});
    \node[left] at (-0.02,{2/3}) {$t$};

    \draw[black,thick] (0,0) rectangle (1,1);

    \draw[decorate,decoration={brace,mirror,amplitude=6pt},thick]
      (0,-0.035) -- ({4*\W},-0.035)
      node[midway,below=8pt] {$\leq t^k$};

  \end{scope}

\end{tikzpicture}%
}
\caption{A pictorial representation of the two parts of the proof. Within the square, each column represents a voter. Each candidate is a block in the column, with height corresponding to their probability mass in $D_t$, ordered by the preference of the voter (higher is more preferred). The block for candidate $a$ sits at height $\rank_v(a;D_t)$ in the column for voter $v$. The left shows the property of \Cref{lem:key}: for each candidate $a$, $\rank_v(a;D_t) > t$ for at most $1 - t$ fraction of voters. The right shows the key property of $S$ (shown as $\{b, c\}$) in the proof of \Cref{thm:five-specific}: at most $t^k$ fraction of voters are not covered by $S$. Every voter that prefers $a$ over $S$ falls into one of these two categories.}\label{fig:two-squares}
\end{figure}

The following key lemma captures the sense in which all candidates tend to be ``ranked low'' by the voters. See \Cref{fig:two-squares} for a visual depiction of what the lemma shows.

\begin{lemmabox}{}{key}
For any $t \in [0, 1]$ there exists a distribution $D_{t}$ over candidates such that for all candidates $a$,
$$\Pr_{v\sim V}[\rank_v(a;D_{t}) > t] \leq 1 - t.$$
\end{lemmabox}

\begin{proof}
Let $\Delta(C)$ denote the simplex of distributions over candidates, and more generally let $\Delta(S)$ denote the simplex of distributions over candidates in the set $S \subseteq C$. Define the functions
$$h^-(a;D) = \Pr_{v\sim V}[\rank_v(a;D) > t] \quad \text{and} \quad h^+(a;D) = \Pr_{v\sim V}[\rank_v(a;D) \geq t].$$
These functions can be thought of as the fraction of voters who rank $a$ ``high'' with respect to $D$ (above the threshold $t$). For intuition, one should think of these as effectively a single function, separated for technical reasons.

Given a distribution $D$ let 
$$M(D) = \{b \in C: h^+(b;D) \geq \max_{a\in C} h^-(a;D)\},$$
roughly representing the subset of candidates which maximize how often they are high with respect to $D$. Now, consider the function $\Psi:\Delta(C)\to 2^{\Delta(C)}$ given by $\Psi(D) = \Delta(M(D))$. With some care, one can show that $\Psi$ satisfies the conditions of the Kakutani fixed-point theorem, but we will defer the details until after the proof to avoid interrupting the flow. The only nontrivial condition to check is that $\Psi$ has a closed graph, which turns out to hold because $h^-(a;D)$ is lower semicontinuous and $h^+(a;D)$ is upper semicontinuous.

It follows that there exists some distribution $D_t$ such that $D_t \in \Delta(M(D_t))$. But then for all candidates $a$, and all candidates $b$ in the support of $D_t$, we have that $h^-(a;D_t) \leq h^+(b;D_t)$. In particular, 
$$h^-(a;D_t) \leq \Ev_{b\sim D_t}[h^+(b;D_t)].$$
Since $h^-(a;D_t)$ is precisely $\Pr_{v\sim V}[\rank_v(a;D_{t}) > t]$, it remains to show that $\Ev_{b\sim D_t}[h^+(b;D_t)] \leq 1 - t$. In fact, this is true even replacing $D_t$ with any distribution $D$ over candidates. 
Intuitively, for each individual voter $v$, $\Pr_{b\sim D}[\rank_v(b;D) \geq t]$ is at most $1 - t$, since the random variable $\rank_v(b;D)$ with $b\sim D$ is stochastically dominated by $r\sim \Unif(0, 1)$, and $\Ev_{b\sim D}[h^+(b;D)]$ is simply the average of $\Pr_{b\sim D}[\rank_v(b;D) \geq t]$ over the voters. More formally, 
\begin{align*}
\Ev_{b\sim D}[h^+(b;D)]
&= \Ev_{b\sim D}\left[\Pr_{v\sim V}[\rank_v(b;D) \geq t]\right] \\
&= \frac1n\sum_{v\in V}\Pr_{b\sim D}[\rank_v(b;D) \geq t] \\
&\leq \frac1n\sum_{v\in V}\Pr_{r\sim \Unif(0, 1)}[r \geq t] \\
&= 1 - t.
\end{align*}
Thus, we can conclude that 
$$\Pr_{v\sim V}[\rank_v(a;D_t) > t] = h^-(a;D_t) \leq \Ev_{b\sim D_t}[h^+(b;D_t)] \leq 1 - t$$
as desired.
\end{proof}

Note that \Cref{lem:key} can be extended\footnote{This extension is similar to \cite[Lemma~1]{DBLP:conf/stoc/CharikarLRV025}, but they have an additional convexity condition on $g$ since they only use the minimax theorem, rather than the stronger Kakutani fixed-point theorem. See \cite{nguyen2026stable} as well for a strengthening of this extension.} to show that for any continuous non-decreasing function $g: [0,1] \to \mathbb{R}_{\geq 0}$, there exists a distribution $D_g$ such that for all candidates $a$, $$\Ev_{v\sim V}[g(\rank_v(a;D_g))] \leq \int_0^1 g(x) \dif x.$$ 
In fact, the proof is a little simpler, since we can just use  $h(a;D) = \Ev_{v\sim V}[g(\rank_v(a;D))]$ in place of $h^+$ and $h^-$, and the continuity of $g$ makes the Kakutani conditions trivial to check. This approach is isomorphic to how the argument of \cite*{DBLP:conf/soda/SongNL26} actually works, with $g$ representing the ``random income distribution.'' They ultimately set $g$ to be a continuous approximation of the function $x \mapsto \textbf{1}[x > t]$ to show a version of \Cref{lem:key} with $1 - t$ replaced with $1 - t + \varepsilon$ for any fixed $\varepsilon > 0$ (which is a practically inconsequential but unnecessary difference).

For the interested reader, we explain exactly how to check that $\Psi$ (defined in the proof of \Cref{lem:key}) satisfies the conditions of the Kakutani fixed-point theorem below.

\paragraph{Checking the Kakutani conditions.} 
As a reminder, here is the statement of the theorem, taken from Wikipedia.  

\begin{theorembox}{(Kakutani fixed-point theorem)}{kakutani}
Let $S$ be a non-empty, compact and convex subset of some Euclidean space $\mathbb{R}^n$. Let $\Phi:S\to 2^S$ be a set-valued function on $S$ such that $\Phi$ has a closed graph, and $\Phi(x)$ is nonempty and convex for all $x \in S$. Then $\Phi$ has a fixed point.
\end{theorembox}

To start, we check the easy conditions. Since $\Delta(C)$ is a simplex, it is clearly non-empty, compact, and convex.  We can also see that $M(D)$ is nonempty for each $D\in \Delta(C)$, since $h^+(a;D)\geq h^-(a;D)$ implies that any candidate $a$ that maximizes  $h^-(a;D)$ is in  $M(D)$. Since $\Delta(M(D))$ is a non-empty simplex, it is convex (and compact as well).

The only condition we have to check carefully is that $\Psi$ has a closed graph, meaning that the set of points $\{(D, Q): Q \in \Psi(D)\}$ is a closed subset of $\Delta(C)\times \Delta(C)$. In other words, for all sequences of points $D^{(i)} \to D$ and $Q^{(i)} \to Q$ such that $Q^{(i)}\in \Psi(D^{(i)})$ for each $i$, we have that $Q \in \Psi(D)$. 

First, we argue that $h^{-}(a;D)$ is \textit{lower semicontinuous}, and  $h^{+}(a;D)$ is \textit{upper semicontinuous}. The characteristic property of a lower semicontinuous function $f$ is that 
$$\liminf_{x\to x_0} f(x) \geq f(x_0),$$
while an upper semicontinuous function $f$ is one that satisfies 
$$\limsup_{x\to x_0} f(x) \leq f(x_0).$$
(These properties are the reason to define $h^-$ and $h^+$ separately.)

It is not hard to check that an indicator function on an open set is lower semicontinuous, and an indicator function on a closed set is upper semicontinuous. (This fact is mentioned on the Wikipedia page for Semicontinuity.)

Since $\rank_v(a; D) = \sum_{b:a \succ_v b}\Pr_D(b)$ is an affine function of $D$, the set $\{D: \rank_v(a;D) > t\}$ is open and $\{D: \rank_v(a;D) \geq t\}$ is closed. It follows that $\textbf{1}[\rank_v(a; D) > t]$ and $\textbf{1}[\rank_v(a; D) \geq t]$ are lower and upper semicontinuous respectively. Since convex combinations of lower/upper semicontinuous functions are also lower/upper semicontinuous, it follows that $h^{-}(a;D)$ and $h^{+}(a;D)$ are lower and upper semicontinuous respectively.

Now we are ready to show that $\Psi$ has a closed graph. Suppose that  $D^{(i)} \to D$ and $Q^{(i)} \to Q$ with $Q^{(i)}\in \Psi(D^{(i)})$ for each $i$. Suppose that $b$ is in the support of $Q$, i.e., $\Pr_Q(b)>0$. Then since $Q^{(i)} \to Q$, we know that for sufficiently large $i$, we must have that $\Pr_{Q^{(i)}}(b)>0$. Since $Q^{(i)} \in \Delta(M(D^{(i)}))$, it follows that for sufficiently large $i$, $b \in M(D^{(i)})$. Unpacking the definition, this means that for all candidates $a \in C$, $h^+(b;D^{(i)}) \geq h^-(a;D^{(i)})$. But then using the characteristic properties of lower semicontinuity and upper semicontinuity, we get 
$$h^+(b;D) \geq \limsup_{i\to\infty} h^+(b;D^{(i)}) \geq \liminf_{i\to\infty} h^-(a;D^{(i)}) \geq h^-(a;D)$$
for all candidates $a\in C$. It follows that $b\in M(D)$, and more generally that every candidate in the support of $Q$ is in $M(D)$. In other words, $Q \in \Delta(M(D)) = \Psi(D)$ as desired. \\

Now we are ready to prove the main theorem.

\begin{theorembox}{\citep*{DBLP:conf/soda/SongNL26}}{five-specific}
In every election, there is a committee $S$ of at most $k$ candidates such that for all candidates $a$, 
$$\tfrac1n|a\succ S| \leq \min_{t\in [0, 1]} (1 - t + t^k) = 1 - \frac{k-1}{k^{k/(k-1)}}.$$
\end{theorembox}

In particular, for $k = 5$, $\min_{t\in [0, 1]} (1 - t + t^k) = 1 - \frac{4}{5^{5/4}}\approx 0.465$, which is less than $\frac12$. It follows that every election has a Condorcet winning set of size at most 5. 

\begin{proof}

Say that a candidate $b$ \emph{covers} voter $v$ if $\displaystyle\min_{a: a\succ_v b} \rank_v(a;D_t) > t$, and a committee $S$ covers $v$ if some candidate $b \in S$ covers $v$ (equivalently, $\displaystyle\min_{a: a\succ_v S} \rank_v(a;D_t) > t$). Using the fact that the random variable $\displaystyle\min_{a: a\succ_v b} \rank_v(a;D_t)$ with $b\sim D_t$ stochastically dominates $r\sim \Unif(0, 1)$, it follows that for each voter $v$, $\displaystyle\Pr_{b\sim D_t}[\text{$b$ covers $v$}] \geq 1 - t$ and $\displaystyle\Pr_{S\sim D_t^k}[\text{$S$ covers $v$}] \geq 1 - t^k$. 

By the probabilistic method, there exists some committee $S$ of size at most $k$ such that the fraction of voters that are \textit{not} covered by $S$ is at most $t^k$. We claim that for this committee, 
$$\max_a \tfrac1n|a \succ S| \leq 1 - t + t^k.$$
Indeed, if a voter $v$ ranks $a$ above $S$, then either $v$ is not covered by $S$ or $\rank_v(a;D_t) > t$. At most $t^k$ fraction of voters can fall into the first category, and at most $1 - t$ fraction of voters can fall into the second (by the property of $D_t$ in \Cref{lem:key}). Choosing $t$ so as to minimize $1 - t + t^k$, the result follows.
\end{proof}

\section{Approximately stable committees}

As a brief addendum, we note that even though \Cref{thm:five-specific} gets strong bounds for small values of $k$, the asymptotics as a function of $k$ are not optimal. In particular, $\min_{t\in [0, 1]} (1 - t + t^k) = \Theta(\frac{\log k}{k})$, but it is possible to get $O(\frac{1}{k})$. 

The key additional idea is to construct the committee recursively. Intuitively, if $k$ is large, rather than sampling the committee in one shot, it can be smarter to start with a few candidates, and then target the rest towards the voters that have low ranks for the current candidates.

\cite*{jiang2020approximately} and \cite{DBLP:conf/stoc/CharikarLRV025} use this recursive approach to give bounds of $\frac{16}{k}$ and $\frac{9.821\ldots}{k}$ respectively, and using the new ideas of \cite*{DBLP:conf/soda/SongNL26}, the bound can be improved to $\frac{4.910\ldots}{k}$. We sketch the proof below.

\begin{theorembox}{\citep*{DBLP:conf/soda/SongNL26}}{apx-stable}
In every election, there is a committee $S$ of at most $k$ candidates such that for all candidates $a$, 
$$\tfrac1n|a\succ S| \leq \frac{4.910\ldots}{k}.$$
\end{theorembox}

In the parlance of \cite*{jiang2020approximately}, the theorem says that $4.91$-stable committees always exist in the ranking setting.

\begin{proof}[Proof sketch.]
Let $\alpha(k)$ be the optimum worst-case bound as a function of $k$. Formally, this is the supremum over elections, of the minimum over committees $S$ of size at most $k$, of $\max_a \tfrac1n |a\succ S|$.

With the convention $\alpha(0) = 1$, we claim that $\alpha(k)$ satisfies the recurrence
$$\alpha(k) \leq \min_{\substack{1\leq \ell \leq k\\t\in [0, 1]}} (1 - t + \alpha(k - \ell)\cdot t^{\ell})$$

Each $\ell$ and $t$ suggests the following way of constructing the committee $S$, following the proof of \Cref{thm:five-specific}. To start, we can choose $\ell$ candidates $T_\ell$ that cover all but $t^\ell$ fraction of voters (by the same argument as before, $T_\ell \sim D_t^\ell$ works with positive probability). Then, recurse on the uncovered voters $U$, finding the committee $T_{k-\ell}$ of size $k - \ell$ that minimizes the fraction of voters in $U$ that rank any candidate $a$ above all the candidates in $T_{k-\ell}$. Finally, the committee $S$ is $T_\ell \cup T_{k - \ell}$. To bound $\tfrac1n|a\succ S|$, we find that the total fraction of voters that can rank a candidate $a$ above $S$ is at most $\alpha(k - \ell)\cdot t^\ell$ from $U$, and at most $1 - t$ from $V\setminus U$, so in total, $\tfrac1n|a\succ S| \leq 1 - t + \alpha(k - \ell)\cdot t^{\ell}$. Choosing the best $\ell$ and $t$, the recurrence follows.

Using the recurrence, we can show that $\sup_k \alpha(k)\cdot k \leq 4.9108$, which implies the theorem. The proof is nontrivial, but largely mechanical, so we omit the details here.
\end{proof}

\paragraph{Acknowledgments.} We thank Th{\`{a}}nh Nguyen for helpful comments and for pointing us to his upcoming paper \citep{nguyen2026stable}.

\bibliographystyle{plainnat}
\bibliography{ref}

\end{document}